\documentclass{PoS}

\PoS{PoS(HEP2005)356}

\title{Fermion Mixing and Soft Leptogenesis in a SUSY $SO(10) \times SU(2)_{F}$ Model}

\ShortTitle{Fermion Mixing and Soft Leptogenesis in a SUSY $SO(10) \times
SU(2)_{F}$ Model}

\author{\speaker{K.T. Mahanthappa}\thanks{Work was partially supported by U.S. Department of Energy under Contract No. DE-FG03-95ER40892.}\\
         Department of Physics, University of Colorado at Boulder, USA\\
         E-mail: \email{ktm@pizero.colorado.edu}}

\author{Mu-Chun Chen\thanks{Fermilab is operated by Universities Research Association Inc. under Contract No. DE-AC02-76CH03000 with the U.S. Department of Energy.}\\
    Theoretical Physics Department, Fermi National Accelerator Laboratory, USA\\
     E-mail: \email{mcchen@fnal.gov}}

\abstract{We have constructed a SUSY $SO(10) \times SU(2)_{F}$ model in which a set of symmetric mass matrices with five texture zeros (having 11 parameters) leads to 22 measurable fermion masses, mixing angles and CP phases, all in agreement with available experimental data within one sigma. The LMA solution for solar neutrinos is obtained as well as neutrino mixing angles and absolute values of neutrino masses. We have investigated the possibility of baryogenesis resulting from soft leptogenesis. We find that, with soft SUSY masses assuming their natural values of the order of a TeV, the observed baryon asymmetry in the Universe can be accommodated in our model. Unlike models with lop-sided textures which give rise to a dangerously large decay rate for $\mu \rightarrow e \; \gamma$, the decay rate we get is much suppressed and yet it is large enough to be accessible to the next generation of experiments. 
\\
\\
COLO-HEP-511, FERMILAB-CONF-05-435-T}

\FullConference{International Europhysics Conference on High Energy Physics\\
                 July 21st - 27th 2005\\     
                 Lisboa, Portugal}

\begin{document}

SO(10) has long been thought to be an attractive candidate for a
grand unified theory (GUT). 
Since a complete quark-lepton symmetry is achieved, it has the promise for
explaining the pattern of fermion masses and mixing. 
Recent atmospheric neutrino oscillation data from
Super-Kamiokande indicates non-zero neutrino masses, giving very
strong support to the viability of SO(10) as a GUT group. Models based on
SO(10) combined with discrete or continuous family symmetry have been
constructed to understand the flavor problem~\cite{paper1,paper2}. 
Most of the models utilize ``lopsided'' mass textures which usually require
more parameters and therefore are less constrained, in addition to giving 
rise to dangerously large rate for $\mu \rightarrow e \gamma$. 
The right-handed neutrino Majorana mass operators in most of these models are made
out of $16_{H}$ which breaks the R-parity at a very
high scale. The aim of this talk, based on 
Ref.~\cite{paper3}, is to present a realistic model based on supersymmetric SO(10) combined
with SU(2) family symmetry which successfully predicts the low energy
fermion masses and mixings. Since we utilize {\it symmetric} mass textures 
(which arise naturally if SO(10) breaks down to SM through the left-right symmetric breaking route) and
$\overline{126}$-dimensional Higgs representations for the right-handed
neutrino Majorana mass operator, our model is more constrained in addition to
having R-parity conserved. SO(10) relates the up-quark mass matrix to 
the Dirac neutrino mass matrix, and the down-quark mass matrix to 
the charged lepton mass matrix. 
 The set mass matrix combination is given by,
at the GUT scale,
\begin{equation}
\scriptstyle{
\label{eq:Mud}
M_{u,\nu_{LR}}=\left(
\begin{array}{ccc}
\scriptstyle{0} & \scriptstyle{0} & \scriptstyle{a} \\
\scriptstyle{0} & \scriptstyle{b}e^{i\theta} & \scriptstyle{c} \\
\scriptstyle{a} & 
\scriptstyle{c} & 
\scriptstyle{1}
\end{array}
\right) d v_{u}, \qquad 
M_{d,e}=\left(
\begin{array}{ccc}
\scriptstyle{0} & \scriptstyle{e} e^{-i\xi} & \scriptstyle{0} \\
\scriptstyle{e} e^{-i\xi} & \scriptstyle{(1,-3) f} & \scriptstyle{0} \\
\scriptstyle{0} & \scriptstyle{0} & \scriptstyle{1}
\end{array}
\right) h v_{d}} \; .
\end{equation}
The smallness of the neutrino masses is explained by 
the type I seesaw mechanism. 
We consider an effective neutrino mass matrix of the form
\begin{equation}\scriptstyle{
\label{eq:Mll}
M_{\nu_{LL}}=M_{\nu_{LR}}^{T} M_{\nu_{RR}}^{-1} M_{\nu_{LR}}
= \left( 
\begin{array}{ccc}  
\scriptstyle{0} & \scriptstyle{0} & \scriptstyle{t} \\
\scriptstyle{0} & \scriptstyle{1} & \scriptstyle{1+t^{n}} \\  
\scriptstyle{t} & \scriptstyle{1+t^{n}} & \scriptstyle{1} 
\end{array} \right) \frac{d^{2}v_{u}^{2}}{M_{R}}}  \; , 
\end{equation}
with $n=1.28$. 
This form is obtained if the right-handed neutrino mass matrix has 
the same texture as that of the
Dirac neutrino mass matrix, 
\begin{equation}\scriptstyle{
\label{eq:Mrr}
M_{\nu_{RR}}=
\left( \begin{array}{ccc}
\scriptstyle{0} & \scriptstyle{0} & \scriptstyle{\delta_{1}} \\
\scriptstyle{0} & \scriptstyle{\delta_{2}} & \scriptstyle{\delta_{3}} \\
\scriptstyle{\delta_{1}} & \scriptstyle{\delta_{3}} & \scriptstyle{1}
\end{array} \right) M_{R}}, \quad \delta_{i}=f_{i}(a,b,c,t,\theta)\;.
\end{equation}
The SU(2) family symmetry is implemented {\it {\'a} la} the
Froggatt-Nielsen mechanism. 
Detailed quantum number assignment, superpotential 
and the VEVs acquired by various scalar fields are given in 
Ref.~\cite{paper3}. 
The input parameters at the GUT scale along with the predictions for charged fermion
 masses and CKM mixing including 2-loop RGE effects are also summarized in Ref.~\cite{paper3}. 
Using  $\Delta m_{atm}^{2}=2.49 \times 10^{-3} \; eV^{2}$ and  
$\Delta m_{\odot}^{2}=7.92 \times 10^{-5} \; eV^{2}$ for the LMA 
solution as input parameters, we determine 
$t = 0.35$ and $M_{R} = 6.57 \times 10^{12} GeV$, which yields 
$\delta_{1}= 0.00120$,
$\delta_{2}=0.000714 e^{i (1.48)}$,
$\delta_{3}=0.0211 e^{-i (0.180)}$. 
The three mass eigenvalues are predicted to be  
$m_{\nu_{1}}=0.0030$ eV,
$m_{\nu_{2}}=0.00939$ eV,
$m_{\nu_{3}}=0.0508$ eV, 
and the mixing angles are predicted to be
$\sin^{2} 2 \theta_{atm} = 1, \;
\tan^{2} \theta_{\odot} = 0.412, \;
\sin^{2}2\theta_{13} = 0.0568$. 
These predictions agree with current 
bounds from experiments within $1 \; \sigma$. 
The strengths of CP violation in the lepton sector are
$(J_{CP}^{l},\alpha_{31},\alpha_{21}) 
= (-0.00983,0.916,-1.52)$, 
and the matrix element for the neutrinoless double 
$\beta$ decay is given by   
$\vert < m > \vert = 3.41 \times 10^{-3}$ eV.

The relevant interactions that 
give rise to lepton flavor violating decays 
come from the soft-SUSY breaking Lagrangian,
\begin{eqnarray}
-\mathcal{L}_{soft} & = &
(m_{\widetilde{L}}^{2})_{ij} 
\widetilde{\ell}_{L_{i}}^{\dagger} \widetilde{\ell}_{L_{j}} 
+ (m_{\widetilde{e}}^{2})_{ij} 
\widetilde{e}_{R_{i}}^{\dagger} \widetilde{e}_{R_{j}} 
+(m_{\widetilde{\nu}}^{2})_{ij} 
\widetilde{\ell}_{R_{i}}^{\dagger} \widetilde{\ell}_{R_{j}} 
+(\widetilde{m}_{h_{d}}^{2})\widetilde{H}_{d}^{\dagger}\widetilde{H}_{d} 
+ (\widetilde{m}_{h_{2}}^{2}) \widetilde{H}_{u}^{\dagger} \widetilde{H}_{u}
\nonumber\\
& &
+ \biggl[ A_{\nu}^{ij} \widetilde{H}_{u} 
\widetilde{\nu}_{R_{i}}^{\ast} \widetilde{\nu}_{L_{j}}
+  A_{e}^{ij} H_{d} \widetilde{e}_{R_{i}}^{\ast} \widetilde{e}_{L_{j}}
+\frac{1}{2} B_{\nu}^{ij} \widetilde{\nu}_{R_{i}} \widetilde{\nu}_{R_{j}}
+ B_{h}  H_{d} H_{u} 
+ h.c. \biggr] .
\end{eqnarray}
As the slepton mass matrix $(m_{\widetilde{L}}^{2})_{ij}$ is 
flavor-blind at the GUT scale with mSUGRA boundary conditions, there is no flavor violation 
at $M_{GUT}$. However, the off diagonal elements in the slepton mass matrix 
can be generated at low energies due to the RG 
evolutions from $M_{GUT}$ to $M_{R}$,  
\begin{equation}
\delta (m_{\widetilde{L}}^{2})_{ij} = -\frac{1}{8\pi} (3m_{0}^{2} + A_{0}^{2}) \times 
\sum_{k=1,2,3} (\mathcal{Y}_{\nu}^{\dagger})_{ik} (\mathcal{Y}_{\nu})_{kj} 
\ln (\frac{M_{GUT}}{M_{R_{k}}}). 
\end{equation}
Here $\mathcal{Y}_{\nu}$ is the Yukawa couplings for the neutrinos in 
the basis where both charged lepton Yukawa matrix and the Majorana 
mass matrix for the RH neutrinos are diagonal. The non-vanishing off-diagonal 
matrix elements in $(\delta m_{\widetilde{L}}^{2})_{ij}$ 
induces lepton flavor violating processes mediated by the superpartners 
of the neutrinos at one-loop, leading to various lepton flavor violating processes. 
The branching ratio for $\ell_{i} \rightarrow \ell_{j} + \gamma$ is given by,
\begin{equation}
Br(\ell_{i} \rightarrow \ell_{j} \gamma) 
= \frac{\alpha^{3}}{G_{F}^{2}m_{S}^{8}} 
\biggl| \frac{-1}{8\pi} (3m_{0}^{2} + A_{0}^{2}) \biggr|^{2} 
\biggr| \sum_{k=1,2,3} ( \mathcal{Y}_{\nu}^{\dagger} )_{ik} (\mathcal{Y}_{\nu})_{kj} 
\ln \biggl(\frac{M_{GUT}}{M_{R_{k}}} \biggr) \biggl|^{2}  \tan^{2} \beta\; .
\end{equation}
Currently the most stringent experimental  bound on the lepton 
flavor violating processes is on the decay $\mu \rightarrow e \gamma$. 
The prediction of our model for  $Br(\mu \rightarrow e \gamma)$ shown in Fig.~(1a) is well 
below the current bound, $1.2 \times 10^{-11}$, yet it is 
within the reach of the next generation of experiments. 

Soft leptogenesis (SFTL) utilizes the soft SUSY breaking sector,  
and the asymmetry in the lepton number is 
generated in the decay of the superpartner of the RH 
neutrinos. It has SUSY breaking as the origin of the CP violation 
and lepton number violation. 
As a result, it allows a much lower bound on 
the mass of the lightest RH neutrino, $M_{1}$, 
compared to that in standard leptogenesis (STDL)~\cite{paper4}. 
The relevant soft SUSY Lagrangian in SFTL that involves lightest RH sneutrinos $\widetilde{\nu}_{R_{1}}$ 
is, 
\begin{equation}
-\mathcal{L}_{soft} = (\frac{1}{2} B M_{1} \widetilde{\nu}_{R_{1}} 
\widetilde{\nu}_{R_{1}} 
+ A \mathcal{Y}_{1i} \widetilde{L}_{i} \widetilde{\nu}_{R_{1}} H_{u} + h.c.) 
+ \widetilde{m}^{2} \widetilde{\nu}_{R_{1}}^{\dagger} 
\widetilde{\nu}_{R_{1}}
\; .\end{equation} 
This soft SUSY Lagrangian together with the superpotential that involves 
the lightest RH neutrino, $N_{1}$, 
$W = M_{1} N_{1} N_{1} + \mathcal{Y}_{1i} L_{i} N_{1} H_{u}$, give rise to the following 
time evolution  Hamiltonian of the  $\widetilde{\nu}_{R_{1}}$-$\widetilde{\nu}^{\dagger}_{R_{1}}$ system,   
\begin{equation}
\frac{d}{dt} \left(
\begin{array}{c}
\widetilde{\nu}_{R_{1}}\\
\widetilde{\nu}_{R_{1}}^{\dagger}
\end{array}\right)
= \mathcal{H}
\left(
\begin{array}{c}
\widetilde{\nu}_{R_{1}}\\
\widetilde{\nu}_{R_{1}}^{\dagger}
\end{array}\right), \quad
\mathcal{H}  =  \mathcal{M} - \frac{i}{2} A, \quad 
\mathcal{M}  =  \left(
\begin{array}{cc}
1 & \frac{B^{\ast}}{2M_{1}}\\
\frac{B}{2M_{1}} & 1
\end{array}\right) \; M_{1}, \quad
A  = \left(
\begin{array}{cc}
1 & \frac{A^{\ast}}{M_{1}}\\
\frac{A}{M_{1}} & 1
\end{array}\right) \Gamma_{1}.
\end{equation} 
The total decay width  $\Gamma_{1}$ of the lightest RH sneutrino is given by, 
$\Gamma_{1} =  \frac{1}{4\pi} 
(\mathcal{Y}_{\nu}\mathcal{Y}_{\nu}^{\dagger})_{11} M_{1}
= 0.374$ GeV. 
The eigenstates of the Hamiltonian are  
$\widetilde{N}_{\pm}^{\prime} = p \widetilde{N} 
\pm q \widetilde{N}^{\dagger}$ and $\biggl( \frac{q}{p} \biggr)^{2} = 
\frac{2\mathcal{M}_{12}^{\ast} - i A_{12}^{\ast}}
{2\mathcal{M}_{12} - i A_{12}} \simeq
1 + Im \biggl( \frac{2\Gamma_{1} A}{BM_{1}} \biggr)$, 
in the limit $A_{12} \ll \mathcal{M}_{12}$.
Similar to the $K^{0}-\overline{K}^{0}$ system, 
the source of CP violation in the lepton number asymmetry here is due to 
the CP violation in the mixing which occurs when the two neutral 
mass eigenstates  
are different from the interaction eigenstates.
Therefore CP violation in mixing is present
as long as 
$\mbox{Im} \biggl( \frac{A\Gamma_{1}}{M_{1} B} \biggr) \ne 0$.  
For this to occur, SUSY breaking, {\it i.e.} non-vanishing $A$ 
{\it and} $B$, is required.
The total lepton number asymmetry integrated over time, $\epsilon$,  
is defined as the ratio of difference to the sum of the decay widths $\Gamma$ 
for $\widetilde{\nu}_{R_{1}}$ and $\widetilde{\nu}_{R_{1}}^{\dagger}$ 
into final states of the slepton doublet $\widetilde{L}$ and the Higgs doublet $H$, 
or the lepton doublet $L$ and the higgsino $\widetilde{H}$ or their conjugates,
\begin{equation}
\epsilon = \frac{\sum_{f=(\widetilde{L}\; H), \; (L \; \widetilde{H})} \int_{0}^{\infty} [
\Gamma(\widetilde{\nu}_{R_{1}}, \widetilde{\nu}_{R_{1}}^{\dagger} \rightarrow
f) - 
\Gamma(\widetilde{\nu}_{R_{1}}, \widetilde{\nu}_{R_{1}}^{\dagger} \rightarrow
\overline{f})]}
{\sum_{f} \int_{0}^{\infty} 
[\Gamma( \widetilde{\nu}_{R_{1}}, \widetilde{\nu}_{R_{1}}^{\dagger} 
\rightarrow f) + 
\Gamma(\widetilde{\nu}_{R_{1}}, \widetilde{\nu}_{R_{1}}^{\dagger}
\rightarrow \overline{f})]}
=\biggl(
\frac{4\Gamma_{1} B}{\Gamma_{1}^{2}+4B^{2}} \biggr)
\frac{Im(A)}{M_{1}} \delta_{B-F}. 
\end{equation}
The final result for the baryon asymmetry is,
\begin{equation}
\frac{n_{B}}{s} 
\simeq -(1.48 \times 10^{-3}) 
\biggl( \frac{Im(A)}{M_{1}} \biggl)
\; R \; \; \kappa,\qquad 
R \equiv \frac{4 \Gamma_{1} B}{\Gamma_{1}^{2} + 4 B^{2}} 
\end{equation}
where the dilution factor $ \kappa =0.00061$  in our model. 
As $\Gamma_{1}$ is of the order of 
$\mathcal{O}(0.1-1) \; GeV$, to satisfy the resonance condition, $\Gamma_{1} = 2|B|$, 
a small value for $B \ll \widetilde{m}$ is thus needed. 
Such a small value of $B$ can be generated  by an operator 
$\int d^{4} \theta ZZ^{\dagger}N_{1}^{2} / M_{pl}^{2}$ in the K\"{a}hler 
potential. In our model, with the parameter $B^\prime  \equiv \sqrt{BM_{1}}$ 
having the size of the natural 
SUSY breaking scale $\sqrt{\widetilde{m}^{2}} \sim \mathcal{O}(1 \; TeV)$, a small 
value for $B$ required by the resonance condition 
$B \sim \Gamma_{1} \sim \mathcal{O}(0.1 \; GeV)$ can thus be obtained. 
The region of the soft SUSY masses that give rise to sufficient baryon asymmetry is shown in Fig.~(1b).
\begin{figure}
\center{
\includegraphics[scale=0.32,angle=270]{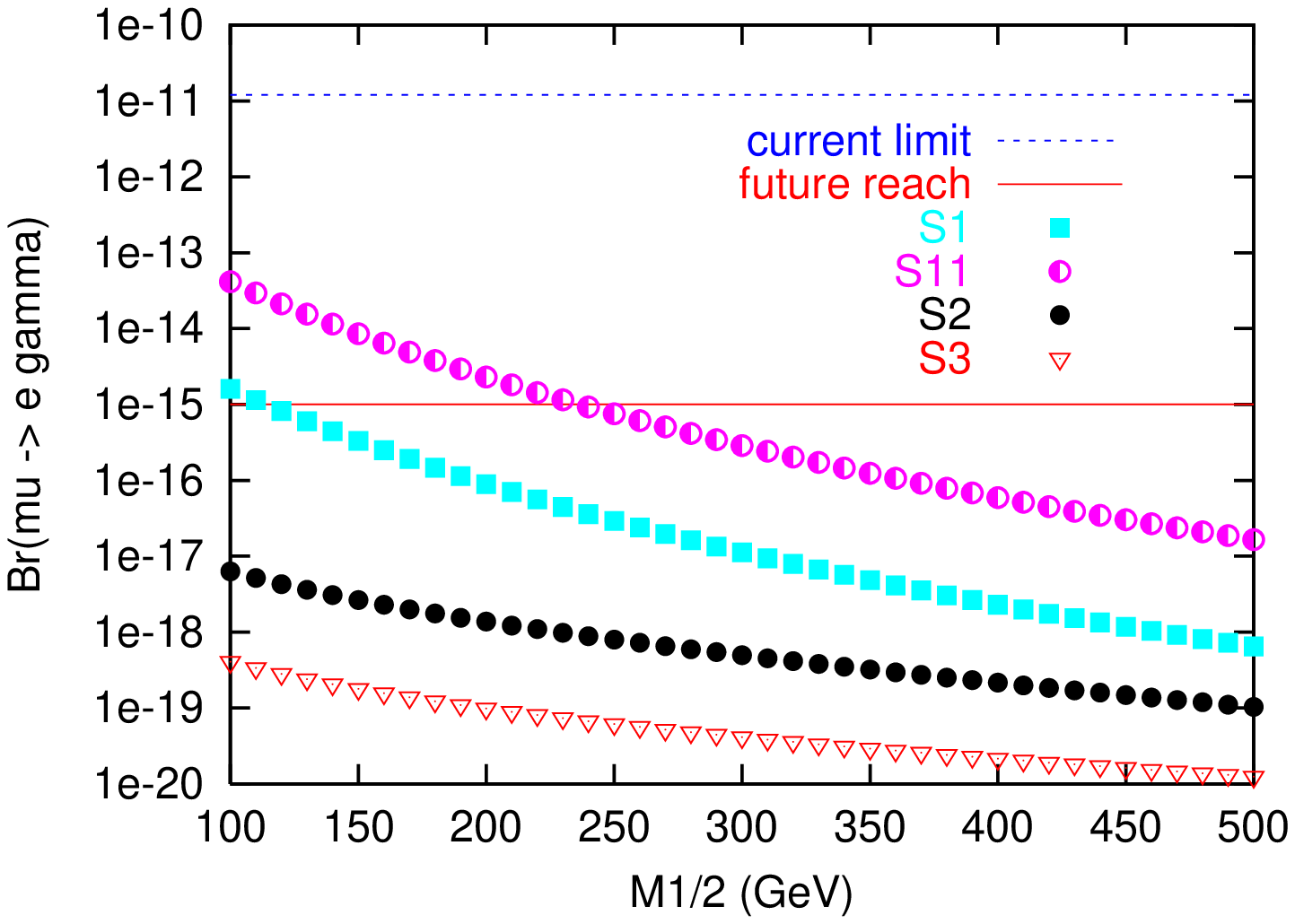}\label{a}
\includegraphics[scale=0.32,angle=270]{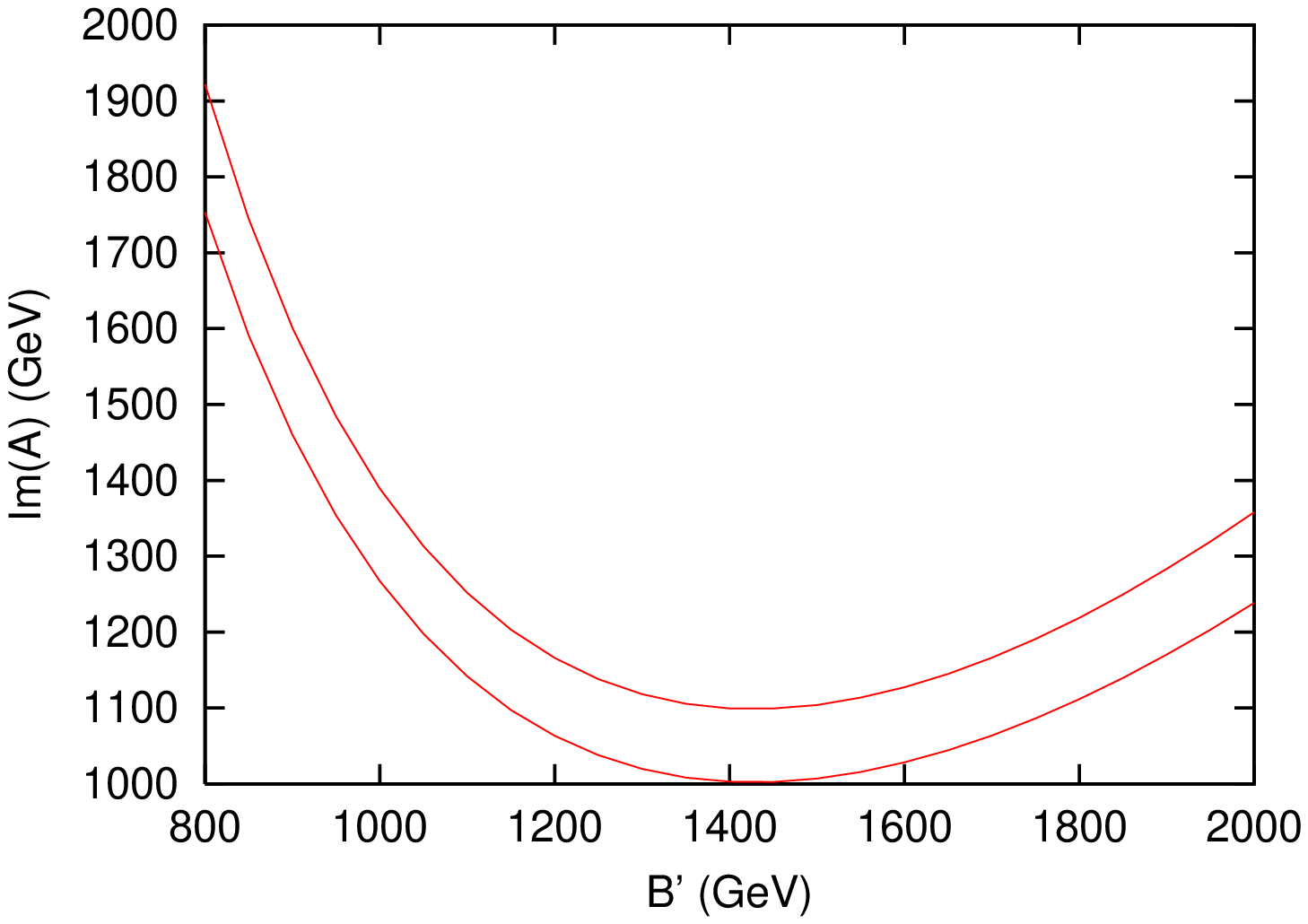}\label{b}}
\caption{\small{(a) Prediction for $\mu \rightarrow e\gamma$ for (S1) $m_{0} = A_{0} = 100$ GeV; 
(S11) $m_{0} = 100$ GeV, $A_{0}=1$ TeV; (S2) $m_0=A_0=500$ GeV; (S3) $m_0=A_0=1$ TeV; 
(b)The region of soft SUSY masses, $A$ and $B^\prime$, that gives rise to observed amount of baryon asymmetry.}}
\end{figure}


\begin{thebibliography}{99}

\bibitem{paper1}
  M.-C.~Chen and K.~T.~Mahanthappa,
  Phys.\ Rev.\ D {\bf 62}, 113007 (2000);  
  {\it ibid} {\bf 65}, 053010 (2002); 
  {\it ibid} {\bf 68}, 017301 (2003);  
  hep-ph/0009059; 
  Int.\ J.\ Mod.\ Phys.\ A {\bf 16}, 3923 (2001); 
  hep-ph/0110037; 
  AIP Conf.\ Proc.\  {\bf 698}, 222 (2004); 
  hep-ph/0409165.


\bibitem{paper2}
  M.-C.~Chen and K.~T.~Mahanthappa,
  Int.\ J.\ Mod.\ Phys.\ A {\bf 18}, 5819 (2003);  
  AIP Conf.\ Proc.\  {\bf 721}, 269 (2004). 
  
\bibitem{paper3}
  M.-C.~Chen and K.~T.~Mahanthappa,
  Phys.\ Rev.\ D {\bf 70}, 113013 (2004). 
  
\bibitem{paper4}
  M.-C.~Chen and K.~T.~Mahanthappa,
  Phys.\ Rev.\ D {\bf 71}, 035001 (2005);  
  M.-C.~Chen,
  {\it ibid} {\bf 71}, 113010 (2005).
  
  

\end{thebibliography}
\end{document}